# Common Artist Music Assistance

Manish Agnihotri, Aditya Rathod, Aditya Jajodia and Chethan Sharma [*]

Department of Information and Communication Technology, Manipal Institute of Technology, MAHE, Manipal, Karnataka, India

{manish.agnihotri, aditya.rathod, aditya.jajodia}@learner.manipal.edu

*chethan.sharma@manipal.edu

**Abstract.** In today's world of growing number of songs, the need of finding apposite music content according to a user's interest is crucial. Furthermore, recommendations suitable to one user may be irrelevant to another. In this paper, we propose a recommendation system for users with common-artist music listening patterns. We use "random walk with restart" algorithm to get relevant recommendations and conduct experiments to find the optimal values of multiple parameters.

## Introduction

The swift increase in the number of songs releasing per day has led to tremendous amount of data being archived online, accessible by making some payment or for free. This leads to situations, where a listener doesn't even get aware of the presence of a song that he could have liked. The escalation of applications like Spotify, has led to various music communities being formed with people listening, sharing, critiquing and even singing and uploading songs giving rise to a huge number of songs, albums, reviewers, artists and listeners. Hence it has become immensely difficult for avid song listeners to find apposite songs. The presence of an efficient recommender system is of utmost importance to tackle this problem.

Recent years have seen the growth of many music based applications like YouTube, Spotify etc. where a user can search for the song of his choice. However, such applications have many downsides like: (i). A user may be unaware of the presence of many songs that he might like. (ii). A user might not himself know, what kind of music he actually likes. (iii).In many cases, it might be difficult to describe one's need in words, especially while searching for something so instinctive like music. Song-listener recommendation methods aim at proposing individualized songs of probable interest for target listeners, hence solving the aforementioned problems.

In this paper, we exploit common artist relation between songs to recommend them to listeners. Additionally, most of the studies neglect the fact that the same recommendation method would not suit all listeners. Hence we have established attributes to categorize listeners suiting our recommendation method, by studying common artist relations in a listener's playlist. So we propose a Common Artist Music Assistance (CAMA), for listeners with artist based song listening patterns.

Our contributions are the following:
- To identify listeners having artist based listening pattern, we have presented 2 features.
- We have proposed a recommendation method, incorporating common-artist-relations among songs, leading to better recommendations to particular target listeners.

## Related Work

Due to the explosion of entertainment content available to users, different recommendation systems are being designed. Three of them are mentioned below.

## Content Based Recommendation

Currently available techniques such as Collaborative Filtering are based on song metadata, and its usage data. Deep Convolutional Neural Networks are being used and they significantly outperform the traditional approaches. They conclude that the advances in Deep Learning translate well to Music Recommendation by bridging the semantic gap and use of latent factor is a feasible solution for recommending new and unpopular music. [1]

**Combination of social media and music content**

The most common approach is the Acoustic based approach, where the audio signal is directly analyzed with respect to frequency, pitch, etc. of the song and the recommendation made is based on the patterns in these.
A hypergraph based approach exploits the information contained in social media and uses hypergraphs to model higher order relation between data (like user-target source), which enables it to combine acoustic based recommendation and collaborative filtering in a unified approach.[2]

**Context aware Music Recommendation System**

Another method uses Bayesian network, fuzzy systems and the utility theory to suggest appropriate music. The Bayesian network takes discrete valued inputs whereas fuzzy systems take inputs in the form of membership values. Context representation is not adequate by discrete values and are hence ignored or rounded off, leading to an error in the training of the model. Fuzzy systems inherit the essence of context and hence lead to a better model. Some continuous attributes are temperature, humidity, noise, etc. Listening to songs about rain on a rainy day or festive songs during festivals is common. Having context about the user's current environment and mood, recommendations systems can be made to work better. [3]

**Problem Statement**

When a listener is interested in a song, he posts it in his playlist and these are used to predict other songs that he might be interested in. In this proposed method, we aim at finding, possibly the most preferred song for a given user.
Fig.1 shows our recommendation scenario, which indicates that a connection between a listener node and a song node represents that user listening to that song, and a connection between a song and a singer indicates that singer singing that song. The first kind of link is used to make recommendations in algorithms like collaborative filtering, whereas the link of second kind is totally neglected. We utilize the second kind of link to make better suggestions by exploiting common singer relations. Listeners search and listen to songs from the same singer after listening to one of his/her songs and we call this "artist-based" song listening pattern. But not all listeners follow this pattern of listening to songs. Some users might search songs based on genre or lyrics or language etc. Hence, we try to solve this problem, by recommending songs to only those listeners who follow the artist-based listening pattern.

**CAMA Recommendation Method**
**Overview**
Our recommendation method is designed by keeping 2 criteria in mind: i) Listeners look for songs associated with same artists. ii) Not all listeners have same patterns. This method is loosely inspired from Xia et.al. 2016 [4]. Hence, we divide our method into 2 modules. The first module is responsible for finding if a listener's patterns are common artist based. The second module recommends songs to a listener based on common artist relations by taking the previous listening patterns as the base. In our first module, we look for ways to extract features from a person's playlist to see if he is a suitable target. In our second module, we use random walk algorithm with restart based on graph to come up with recommendations. We then select top-k recommendations as our result. Random walk algorithm is an algorithm used in recommender systems [5].

**Module 1: Target Listener's Selection**

Our song recommendation method is suitable to only those users who have a pattern of listening to songs on a common artist basis. So we design a way to identify such listeners first. We have proposed two attributes, which mathematically exploit the common artist relation between songs and help identify such listeners.

- Cama1: the ratio of the number of pair-wise songs having one or more same artists, to the number of all pairwise songs for a listener.
- Cama2: the ratio of number of times the most frequently occurring artist has occurred in a listener's playlist to the total number of songs in the listener's playlist

For a listener, if these scores are larger than a given threshold, this listener is considered to be listening to songs on a common artist basis and is selected as our target for recommendation using CAMA. We use Fig.2 to explain the computation process of Cama1 and Cama2.

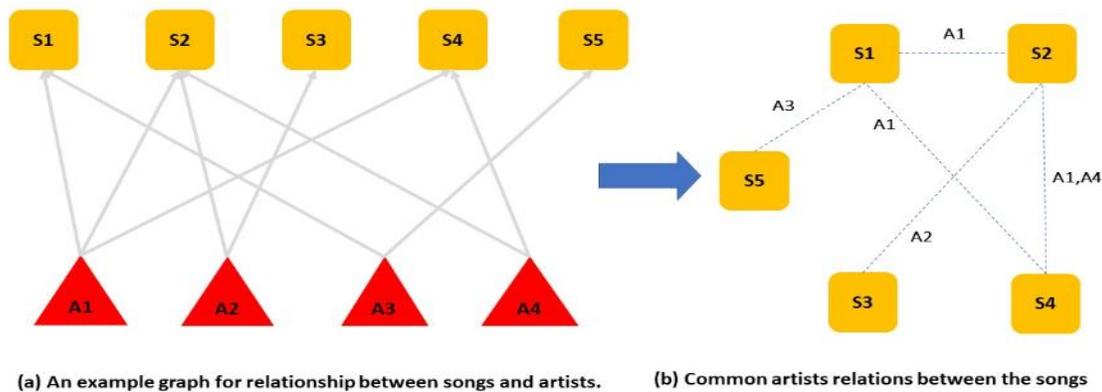

Fig. 1: *Example scenario for a listener*

Fig. 1a shows the composing relation between an artist and a song for particular target listener L, wherein every edge connects a song to one of its artists. For instance, song S1 is connected to its two artists A1 and A3 through two distinct edges. We take two songs to be related if they are connected to the same artist in Fig. 1a. Following this method, we transform this, into another graph indicating common-artist-relation, represented in Fig. 1b, wherein any two songs are connected in case they have one or more common artist(s). This new figure helps us acquire the number of songs taken two at a time with common artist(s) and it equals 5. In this case, the total number of combinations of all songs taken two at a time is equal to $^5C_2=10$.

Consecutively, Cama1 equals 5/10=0.5. Additionally, artist A1 appears the maximum number of times(thrice) for this listener. Hence, Cama2 equals 3/5=0.6. Considering the minimum threshold values of Cama1, Cama2 are fixed at 0.4 and 0.5, respectively, then 0.5>0.4 and 0.6>0.5 Hence, this listener suits our CAMA method of recommendation. A listener crossing only Cama1 and not Cama2 indicates that this listener has listened to songs with a very small number of songs for any common artist. For instance, if a user has 10 songs in his playlist and the common artist relation follows the pattern: Songs 1 and 2 have a common artist, 2 and 3 have another common artist, 3 and 4 have some other common artist and so on. Here our Cama1 score becomes 9/(10 C 2) = 0.2, which might be above our decided threshold but this listener is clearly not a potential one for our algorithm as all common artists have only 2 songs in his playlist. Such a situation might also occur arbitrarily, since we consider all the artists in building our relation.

A listener crossing only Cama2 and not Cama1 indicates that this listener has listened to many songs of a very small number of artists but his whole playlist doesn't represent the same pattern. For instance a listener having 50 songs in his playlist with 25 songs having a common artist and the other 25 with no common artist brings up the Cama2 score to as high as 25/50 = 0.5. But again, this is clearly not a potential target listener for us. Such a situation indicates that the listener has a favorite artist to whom he listens to, quite often, but in general doesn't follow a common artist approach for listening to songs.
In the first case the listener doesn't cross the threshold set for Cama2 due to a very small number of songs for any common artist, and in the second case, the listener doesn't cross the Cama1 threshold due to the whole playlist not following the common artist trend. Thus, only a combined way of considering both these scores to select our target listeners is an ideal approach.

Other features to decide the relevance of a listener for our method can be explored in our future work. For this paper we take only these two features and plot graphs after conducting experiments to substantiate their potency in Experiments Section.

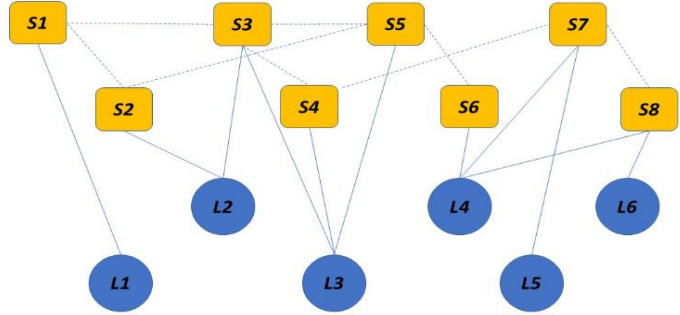

Fig. 2: *Example graph for song ranking*

**Module 2: Generating Song Recommendations**

**Constructing Graph:** In the lack of a dataset which captures user's listening history and details on songs, we generated our own dataset. Our dataset contains two files. The first contains each user's playlist and the second contains artist-list of each song. We generated the dataset for 100 users, each having a playlist of more than 10 and less than 20 songs. We used 50 different songs with each song having more than 1 and less than 5 artists in our second file. The reason for having songs with more than one artist was to give equal importance to all the artists of a song, We ensured having a minimum of 10 songs in every user's history because our method in Module 1 evaluates whether a listener is our target listener and we need at least 10 songs to predict that efficiently. It also ensures at least 2 songs per user in our test dataset.

This dataset is represented in form of a graph to implement random walk with restart on it. Our graph contains 2 types of node, namely listener node and song node. Listener node L = {$L_1$, $L_2$, …. $L_n$} is connected to all song nodes that it listens to, and two song nodes are connected if they share a common artist.

**Transition Probability Computation:**

Random walk defines a shift from a node to another node. Elements of matrix aij represents transition probability from node i to node j. The mathematics to obtain it are defined as follows. All neighbor nodes have equal values. Since there is no direct connection between any two listeners, the transition probability is

$$TP_{LL}(i-j) = 0 \qquad (1)$$

The transition-probability of moving from listener to a song is

$$TP_{LS}(i,j) = \frac{W_{LS}(i,j)}{\sum_k W_{LS}(i,k)} \qquad (2)$$

Since a song is connected with both another song and listener there are two transition probability defined as

$$TP_{SS}(i,j) = \frac{W_{SS}(i,j)}{\sum_{k1} W_{SL}(i,k1) + \sum_{k2} W_{SS}(i,k2)} \qquad (3)$$

and

$$TP_{SL}(i,j) = \frac{W_{SL}(i,j)}{\sum_{k1} W_{SL}(i,k1) + \sum_{k2} W_{SS}(i,k2)} \qquad (4)$$

The transition probability matrix is

$$TP = \begin{bmatrix} TP_{LL} & TP_{LS} \\ TP_{SL} & TP_{SS} \end{bmatrix} \qquad (5)$$

**Random Walk with Restart:**

The output of the algorithm is ranking of songs for different listeners. Ranking of the songs which are on his playlist will be high, but for the obvious reasons, we will not take those rankings into consideration. We use these songs to find another song following the path, listener →song → song. Also to reach another song, we have one more path which is, listener → song → listener → song. Using these two paths, we can incorporate features of common artists' relationship and collaborative filtering.

The algorithm starts with the current listener $n_0$ in consideration. Then using random walk with restart, we move to another node(x) with probability alpha, having link transition probability TP(n0,nx) and then from nx to ny and this continues. Top N songs with the highest ranking, not present in the listener's playlist, will be recommended.

**Algorithm 1.** Graph-based music ranking.

**Input:**
    Graph, *G*;
    Transition probability matrix, *TP*;
    Target listener node, $n_0$;
    Random walk probability, $\alpha$;
    Maximum step length of iteration, *maximum_step*;

**Output:**
    Ranking scores of all song nodes, *ScoreSong (1:m)*;
    // m song nodes

1: Define ranking scores of all nodes, *ScoreAll(1 : n+m)* ; // n+m nodes
2: **for** each $n \in N_L \cup N_S$ **do**
3:     *ScoreAll(n)* = 0; //initial ranking scores are 0
4: **end for**
5: *ScoreAll($n_0$)* = 1;
6: **for** *step* = 0; *step* < *maximum_step*; *step* ++ **do**
7:     **for** each $n \in N_L \cup N_S$ **do**
8:         *tempScore(n)* = 0; //initial values are 0
9:     **end for**
10:     **for** each $n_x \in N_L \cup N_S$ **do**
11:         **for** each $n_y \in N_L \cup N_S$ **do**
12:             *tempScore($n_y$)* = $\alpha$ × *ScoreAll($n_x$)* × *TP($n_x,n_y$)* + *tempScore($n_y$)*;
13:         **end for**
14:         **if** $n_x == n_0$ **then**
15:             *tempScore($n_x$)* = *tempScore($n_x$)* + 1 - $\alpha$;
16:         **end if**
17:     **end for**
18:     *ScoreAll* = *tempScore*;
19: **end for**
20: *ScoreSong (1 : m)* = *ScoreAll(n+1: n+m)*;
    // select ranking scores of song nodes
21: **return** *ScoreSong (1 : m)*;

## Experiments

We divided our data into train and test sets in the ratio 4:1 using the following procedures. We divided the song history from dataset file into train and test, for every user. Training set is considered as the songs, listener has listened to, and based on that, we make recommendations. Performance is evaluated by checking the presence of recommended songs, in the test set. Three performance metrics are used to verify our recommendation quality: (i) Precision, (ii) Recall and (iii) F1, that are used universally and extensively in scientific literature in the field of information retrieval science and recommender systems.

**Influence of Walking Probability**

Walking probability represents the probability of walk between the vertexes and its neighbors, whereas (1-alpha) is the probability to walk back to the source node. Alpha is the weight we give to all the connected nodes to a vertex and has an impact on the recommendation quality. In our experiments, we varied the value of alpha from 0.2 to 0.8 with a step of 0.2. Alpha's value of 0.8 gave the best result and hence we chose it as the optimal value for further experiments.

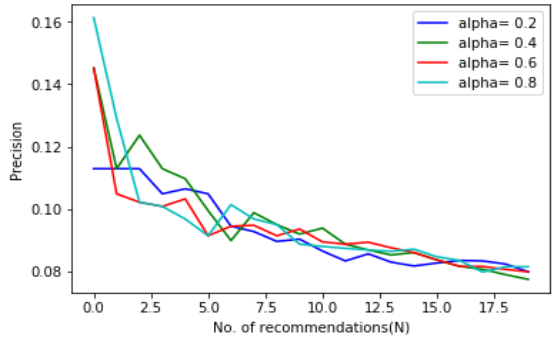

Fig. 3: *Precision of CAMA for different values of alpha*

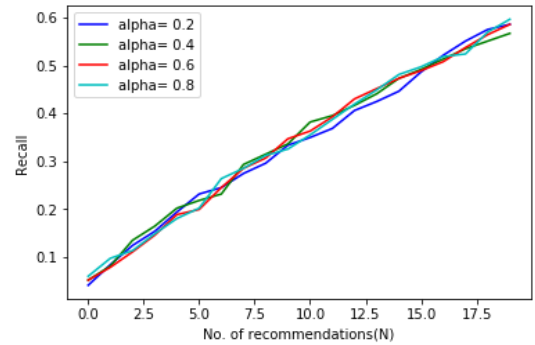

Fig. 4: *Recall of CAMA for different values of alpha*

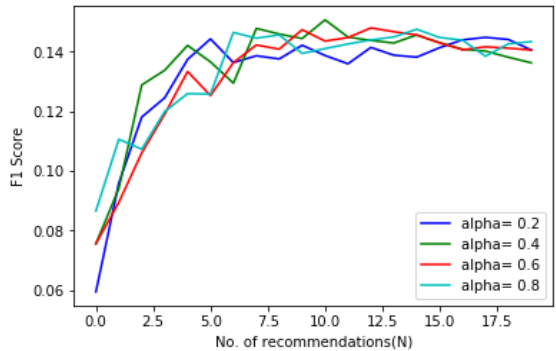

Fig. 5: *F1 score of CAMA for different values of alpha*

**Results**

Due to the absence of a dataset which suited the needs of our algorithm, we had to generate our own dataset. The random songs generated in the playlists follow a normal distribution. The reason for choosing normal distribution was to replicate a real world scenario. A uniform distribution would have indicated that all songs are being similarly liked by the listeners which is different from the real world scenario, where hit songs are heard by more people and vice-versa. The artists of each song have also been randomly generated with a normal distribution, keeping in mind that not all artists sing even approximately same number of songs.

Subsequently, we apply the algorithm on the generated dataset to measure the performance based on metrics: Precision, Recall and F1 score as mentioned earlier. We plot the values of these metrics against the number of recommendations. We have three hyper-parameters: alpha and the threshold values of

Cama1 and Cama2 to get the optimal values for. Hence we get 9 graphs, by plotting each metric on varying all three hyper-parameters independently. These graphs are illustrated in figures 3 to 11. The first three plots, for varying alpha, shows its best value to be 0.8.

The next three plots, for varying threshold values of Cama1 Score, show that our recommendation becomes better as this value is increased, thus backing our assumption that a listener with higher Cama1 score is more likely to suit our recommendation method. But this value cannot be exceeded after a limit, as the number of listeners crossing that threshold will decrease, thus reducing the reach of our recommendation method.

The plots of Cama2 score follow a similar trend, and thus have the same implications as stated above for Cama1 score.

The jitter in the plots is due to training being done on a small data. But the trend of these metrics with respect to our scores is clearly evident in these plots and will remain same even on a larger dataset.

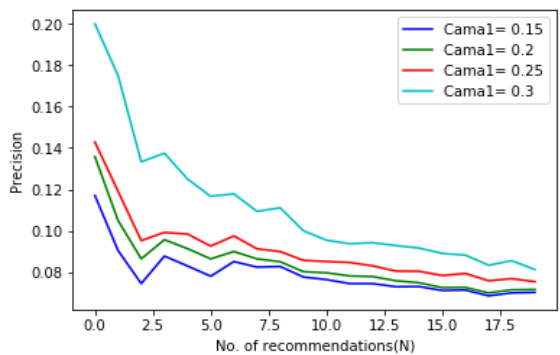

Fig. 6: *Precision of CAMA for different values of cama1*

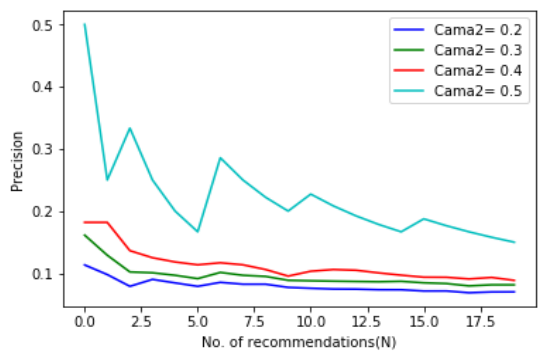

Fig. 7: *Precision of CAMA for different values of cama2*

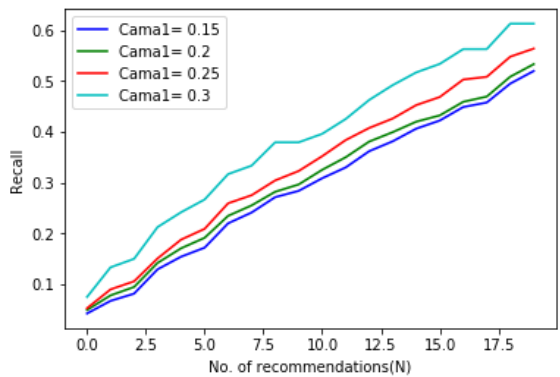

Fig. 8: *Recall of CAMA for different values of cama1*

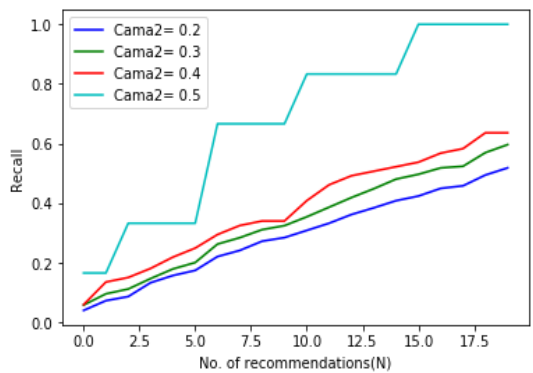

Fig. 9: *Recall of CAMA for different values of cama2*

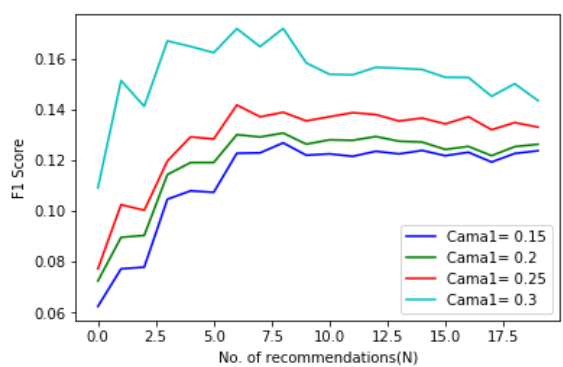 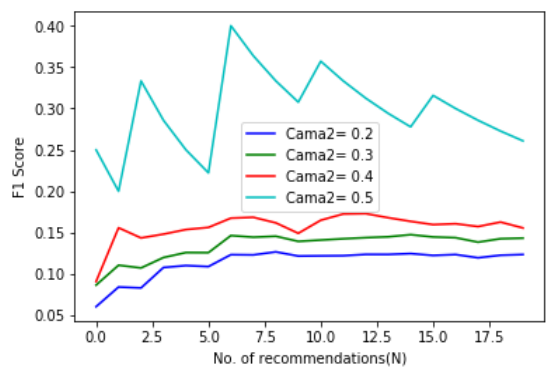

Fig. 10: *F1 of CAMA for different values of cama1*     Fig. 11: *F1 of CAMA for different values of cama2*

**Conclusion**

In this paper, a way of finding recommendations for listeners has been proposed. We are using the hypothesis that listeners tend to listen to songs of artists who have the same genre or style. Before recommending we make sure that the target listener passes two scores namely, Cama1 and Cama2. Using the data from existing listeners and their playlist we build a graph-based song ranking algorithm. After varying different parameters, we found that giving five recommendation, gave us best result. Due to limitations in data size and computational resources, we didn't test our algorithm on large scale data. We are very confident that even on large-scale data, our algorithm will give similar, if not better results.

In future, we plan to add new scores which allow us to recommend songs to more users, at the same time do not degrade the quality of recommendation. We also plan to tag along composer, music directors, etc. with singers, which would add another dimensionality to the data set.

**References**

[1] A.V.D. Oord, S. Dieleman, B. Schrauwen, Deep content-based music recommendation, Advances in Neural Information Processing Systems (NIPS), 2013

[2] Jiajun Bu, Shulong Tan, Chun Chen ," Music recommendation by unified hypergraph: combining social media information and music content ", Proceedings of the 18th ACM international conference on Multimedia Firenze, Italy 2010, Pages 391-400.

[3] H.C. Chen, A.L.P Chen , A Context-Aware Music Recommendation System Using Fuzzy Bayesian Networks with Utility Theory, International Conference on Fuzzy Systems and Knowledge Discovery (FSKD), 2006

[4]  F. Xia, H. Liu, I. Lee, L. Cao, Scientific Article Recommendation: Exploiting Common Author Relations and Historical Preferences, IEEE Transactions on Big Data, 2016.

[5] M. Gori, A. Pucci, Research Paper Recommender Systems: A Random-Walk Based Approach, IEEE/WIC/ACM International Conference on Web Intelligence, 2006